\documentclass[%
reprint,
superscriptaddress,
amsmath,amssymb,
aps,
floatfix,
prl
]{revtex4-2}

\usepackage{graphicx}% Include figure files
\usepackage{dcolumn}% Align table columns on decimal point
\usepackage{multirow}% Multi-row cells in tables
\usepackage{bm}% bold math
\usepackage{hyperref}% add hypertext capabilities

\begin{document}

% Custom macros for this paper
\newcommand{\phos}{$^{26}\mathrm{P}$}
\newcommand{\sili}{$^{26}\mathrm{Si}$}
\newcommand{\alum}{$^{25}\mathrm{Al}$}
\newcommand{\magn}{$^{24}\mathrm{Mg}$}

\preprint{APS/123-QED}

\title{Resolving the isospin-mixing anomaly in $^{26}$Si from $^{26}$P $\beta$-decay}
 \author{B.\,S.\,O.~Johansson}
 \affiliation{Institut for Fysik \& Astronomi, Aarhus Universitet, Aarhus, 8000, Denmark}
\author{E.\,A.\,M.~Jensen}
 \email[Contact author: ]{erik.jensen@chalmers.se}
 \affiliation{Institutionen för fysik och astronomi, Chalmers tekniska högskola, 412 96, Göteborg, Sweden}
 \author{J.\,S.~Nielsen}
 \affiliation{Institut for Fysik \& Astronomi, Aarhus Universitet, Aarhus, 8000, Denmark}
\author{A.~Adams}
 \affiliation{Department of Physics and Astronomy, Michigan State University, East Lansing, Michigan 48824, USA}
 \affiliation{Facility for Rare Isotope Beams, Michigan State University, East Lansing, Michigan 48824, USA}
\author{M.\,J.\,G.~Borge}
 \affiliation{Instituto de Estructura de la Materia, CSIC, Madrid, 28006, Spain}
\author{J.~Dopfer}
 \affiliation{Department of Physics and Astronomy, Michigan State University, East Lansing, Michigan 48824, USA}
 \affiliation{Facility for Rare Isotope Beams, Michigan State University, East Lansing, Michigan 48824, USA}
\author{H.\,O.\,U.~Fynbo}
 \affiliation{Institut for Fysik \& Astronomi, Aarhus Universitet, Aarhus, 8000, Denmark}
\author{B.~Jonson}
 \affiliation{Institutionen för fysik och astronomi, Chalmers tekniska högskola, 412 96, Göteborg, Sweden}
 \author{M.\,V.~Kristensen}
 \affiliation{Institut for Fysik \& Astronomi, Aarhus Universitet, Aarhus, 8000, Denmark}
\author{M.~Madurga}
 \affiliation{Department of Physics and Astronomy, University of Tennessee, Knoxville, Tennessee 37996, USA}
 \author{E.\,D.~Nielsen}
 \affiliation{Institut for Fysik \& Astronomi, Aarhus Universitet, Aarhus, 8000, Denmark}
\author{K.~Riisager}
 \affiliation{Institut for Fysik \& Astronomi, Aarhus Universitet, Aarhus, 8000, Denmark}
\author{C.\,S.~Sumithrarachchi}
 \affiliation{Facility for Rare Isotope Beams, Michigan State University, East Lansing, Michigan 48824, USA}
\author{L.\,J.~Sun}
 \affiliation{Facility for Rare Isotope Beams, Michigan State University, East Lansing, Michigan 48824, USA}
\author{O.~Tengblad}
 \affiliation{Instituto de Estructura de la Materia, CSIC, Madrid, 28006, Spain}
\author{L.\,E.~Weghorn}
 \affiliation{Department of Physics and Astronomy, Michigan State University, East Lansing, Michigan 48824, USA}
 \affiliation{Facility for Rare Isotope Beams, Michigan State University, East Lansing, Michigan 48824, USA}
\author{T.~Wheeler}
 \affiliation{Department of Physics and Astronomy, Michigan State University, East Lansing, Michigan 48824, USA}
 \affiliation{Facility for Rare Isotope Beams, Michigan State University, East Lansing, Michigan 48824, USA}
\author{C.~Wrede}
 \affiliation{Department of Physics and Astronomy, Michigan State University, East Lansing, Michigan 48824, USA}
 \affiliation{Facility for Rare Isotope Beams, Michigan State University, East Lansing, Michigan 48824, USA}

\date{\today}

\begin{abstract}
The $\beta$-delayed one- and two-proton emission of the proton-dripline nucleus $^{26}$P provides a sensitive probe of isospin symmetry-breaking.
A recent study reported the observation of a strongly isospin-mixed doublet in $^{26}$Si, comprised of the $T = 2$ Isobaric Analog State (IAS) and a state at 13380~keV, yielding a record-high isospin mixing matrix element.
We report a new high-resolution measurement of the $^{26}$P $\beta$-delayed charged-particle decay performed at the Facility for Rare Isotope Beams (FRIB), shedding new light on the isospin mixing in this decay.
By utilizing low-energy beam implantation from the Advanced Cryogenic Gas Stopper (ACGS), we obtained clean one-proton and two-proton decay spectra.
Our high statistics study show discrepancies in decay branches and the existence of a 13380~keV state in $^{26}$Si is unsupported by our data.
Our deduced branching ratio to the IAS is fully consistent with the expectation from the Fermi sum rule.
Consequently, the claim of strong isospin mixing in $^{26}$Si is refuted, reinstating standard theoretical expectations for isospin mixing in nuclei.
\end{abstract}

\maketitle

Isospin symmetry, which arises from the charge independence of the strong nuclear force, dictates that the superallowed Fermi $\beta$-decay strength should be concentrated in a single transition to the Isobaric Analog State (IAS). This is because the IAS, under exact isospin symmetry, has the same nuclear wave function as the parent nucleus with a single proton replaced by a neutron, which maximizes the spatial overlap. The Coulomb interaction and charge-dependent components of the nuclear force can break this symmetry slightly and give rise to fragmentation of the Fermi strength where the IAS mixes with closely lying states of the same spin and parity, $J^\pi$. The observation of such isospin-mixed states allows for precision tests of isospin symmetry-breaking in nuclear structure models.

Representative cases of such mixing are known across the chart of nuclides. In the lightest systems, such as the decay of $^{8}$B to $^{8}$Be and $^{12}$N to $^{12}$C, the mixing strength is typically inferred from a combination of Fermi strength fragmentation and isospin-forbidden particle emission widths, analyzed within an $R$-matrix or cluster-model framework \cite{FERNANDEZRUIZ2026140791, Barker1969, Tilley2004, Tilley1993}. In heavier systems, several clean cases of IAS mixing have also been established in $\beta$-decay studies, where the mixing manifests as a strongly $\beta$-fed doublet in the charged particle or $\gamma$-ray spectrum. The higher level density in heavier nuclei increases the probability to find a $J^\pi$-matched state close in energy with the IAS, making such doublets more accessible experimentally. Notable examples include the decays of $^{31}$Cl, $^{32}$Ar, $^{37}$Ca, $^{55}$Cu, and $^{56}$Zn~\cite{Bennett2016, Bhattacharya2008, Trinder1995, Tripathi2013, Orrigo2014}.

While the IAS can be populated in the $\beta$-decay of all nuclei with more protons than neutrons, nuclei at or close to the proton-drip line, occupy a privileged position for isospin-symmetry tests (for instance the recent review in \cite{Pfutzner2023}): The large $Q$-value for electron capture, $Q_\text{EC}$, places the IAS at high excitation energy in the daughter. Since this excitation energy typically exceeds the nucleon-emission threshold, $\beta$-delayed one- and two-proton ($\beta 1p$/$\beta 2p$) emission channels provide probes of the superallowed Fermi strength that are inaccessible in near-stable systems. Precise measurements in such nuclei directly benchmark the isospin symmetry-breaking corrections, parameterized as $\delta_C$, relevant for the extraction of $|V_{ud}|$ from superallowed $\beta$-decays~\cite{Hardy2020,Shukla2026}. The isospin symmetry-breaking correction framework has moreover been extended from superallowed Fermi to Gamow-Teller transitions, where Coulomb-induced isospin mixing ($\delta_{C1}$) and radial wave-function mismatch ($\delta_{C2}$) generate measurable mirror asymmetries between conjugate $\beta^+$ and $\beta^-$ decays~\cite{Xayavong2025}.

\begin{figure}[t]
\centering
\includegraphics[width=0.95\columnwidth]{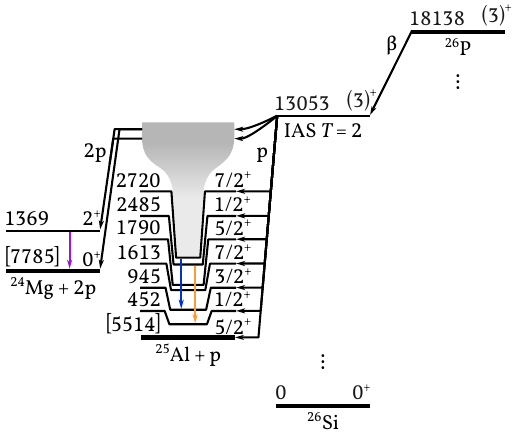}
\caption{\label{fig:decayscheme}%
Partial decay scheme for the $\beta$-delayed one- and two-proton emission of \phos{}.
Two-proton emission proceeds primarily via higher excited states in \alum{}, indicated by the shaded gray shape.
Energies are given in keV relative to the ground state of \sili{}, except for excited states in \alum{} and \magn{} which are given relative to their respective ground states.
The mass of \phos{} is adopted from \cite{Chen2025}, and the IAS energy is derived from our measured $\beta$-delayed two-proton spectrum.
Other level parameters are adopted from \cite{AME2020,Basunia2025}.
The $\gamma$-rays indicated are used in Figs. \ref{fig:q1p} and \ref{fig:q2p}.}
\end{figure}

The proton dripline isotope $^{26}$P undergoes a $\beta$-decay of large $Q$-value towards excited states in $^{26}$Si many of which are unbound towards one- and two-proton emission, ultimately feeding states in $^{25}$Al and $^{24}$Mg; see Fig. \ref{fig:decayscheme}. The \phos{} decay has been studied in experiments at Berkeley \cite{Honkanen1983, Cable1984}, GANIL \cite{Thomas2004}, NSCL \cite{Bennett2013, PerezLoureiro2016}, Dubna \cite{Janiak2017}, and HIRFL \cite{Liang2020,Liu2022}. For this isobaric multiplet in particular, the proton-halo character of \phos{} --- in which the weakly-bound valence protons acquire a diffuse spatial distribution --- was suggested by P\'erez-Loureiro \textit{et al.}~\cite{PerezLoureiro2016} to be the cause of significant mirror asymmetry. Subsequent shell-model calculations by Kaneko~\textit{et al.}~\cite{Kaneko2019} investigated this interpretation in further theoretical detail, predicting a large mirror asymmetry in the Gamow-Teller decay strength amplified by the enhanced $\delta_{C2}$ and heightened sensitivity to Coulomb effects.

A recent study by Liu \textit{et al.}~\cite{Liu2022} reported a strongly isospin-mixed doublet in $^{26}$Si in the decay of \phos{}. Excited states of \sili{} at these high energies ($\gtrsim 10$~MeV) are accessible exclusively through \phos{} $\beta$-delayed particle emission, as they lie beyond the reach of conventional transfer reactions. Based on $\beta 1p$ and $\beta 2p$ measurements, the authors assigned the IAS to a state at 13055~keV and proposed two new, highly excited states at 11912~keV and 13380~keV, neither of which had been observed in any prior experiment. The latter was identified as a Fermi-transition-dominated partner to the IAS, yielding an extracted isospin mixing matrix element --- the off-diagonal Hamiltonian coupling between the two states~\cite{Liu2022} --- of $130(21)$~keV.  In the \emph{sd}-shell, this value represents the strongest isospin mixing at the highest excitation energy, and across the largest level spacing ($\Delta E = 325$~keV), ever reported in a $\beta$-decay experiment, challenging standard theoretical descriptions of isospin mixing in the $sd$-shell \cite{Brown2006}.

In the 2022 study by Liu \textit{et al.}~\cite{Liu2022}, $^{26}$P ions were produced via projectile fragmentation and implanted into a stack of silicon detectors at high kinetic energies \mbox{($\sim 80$~MeV/u)}. Detecting $\beta 2p$ emission under these conditions requires reconstructing the total two-proton decay energy, $Q_{2p}$, by tracking individual protons as they escape the implantation zone and enter adjacent detectors. This procedure is  susceptible to $\beta$-particle energy summing, which degrades the energy resolution and can distort spectral shapes. Furthermore, impurities in the secondary beam can introduce backgrounds. The $Q_{2p}$ spectrum in Ref.~\cite{Liu2022} used to justify the 11912~keV and 13380~keV states is characterized by low statistics spread across a baseline containing $1p$ decays and contaminant backgrounds. Under such conditions, unresolved $1p$ artifacts or statistical fluctuations can be misidentified as genuine $2p$ resonances, rendering the extracted branching ratios and $\log ft$ values susceptible to systematic errors. These limitations call for a measurement with improved energy resolution, suppressed backgrounds, and higher statistics.

In this Letter, we resolve the structure of the Fermi strength in $^{26}$Si populated by $^{26}$P $\beta$-decay. The present measurement was performed during the first FRIB/ACGS campaign at the Facility for Rare Isotope Beams (FRIB), utilizing the experimental setup and beam delivery methods detailed in our recent studies of $^{25}$Si \cite{Jensen2024Si25} and $^{22}$Al \cite{Jensen2026}.
The experiment conducted at FRIB was with a 5~kW primary beam of $^{36}$Ar which was accelerated to 210~MeV/u before impinging on an 8.07~mm carbon production target. The resulting in-flight cocktail beam was momentum-to-charge-separated in the Advanced Rare Isotope Separator (ARIS) \cite{Portillo2023, Fukushima2023}, reducing the beam energy to 106~MeV/u. Subsequently, the beam was guided to the Gas Stopping Area for momentum compression, thermalization, and purification. A low-energy 30~keV beam of $^{26}$P was extracted from the Advanced Cryogenic Gas Stopper (ACGS) \cite{Lund2020, Ringle2021} and implanted into a thin carbon foil. Mass selection via an analyzing magnet supressed contaminants, although a sizeable contamination of $^{25}$SiH on the order of 40\,\% of the beam passed through at the same mass-to-charge ratio $A/q$. In a separate part of the experiment a measurement of a pure beam of $^{25}$Si made it possible to separate the $1p$ peaks of \phos{} from $^{25}$Si~\cite{Jensen2024Si25}.

The catcher foil, from which the stopped radioactive beam decayed, was surrounded by a compact array of segmented silicon detector telescopes, allowing for highly granular, background-suppressed charged-particle detection \cite{Jensen2024Si25}. The detectors were placed in $\Delta E$--$E$ telescope configurations, with the $\Delta E$ detectors being 16$\times$16 Double-sided Silicon Strip Detectors (DSSDs) and the $E$ detectors being 1-channel Single-sided Silicon Detectors (SSDs). Outside the vacuum chamber two High Purity Germanium detectors were placed for detecting $\gamma$-rays in coincidence with protons measured in the Silicon detectors.

Approximately $4\times10^6$ low-energy $^{26}$P ions were stopped in the catcher foil. We detected 13,400 $\beta 2p$ events and 470,000 $\beta p$ events, representing a significant increase in statistics compared to previous measurements. The implantation rate at the setup is estimated to be around 35--40 ions per second based on the number of observed $\beta$-delayed protons throughout the experiment, correcting for known branching ratios \cite{Thomas2004} and the solid angle coverage of the array.

\begin{figure}[t]
\centering
\includegraphics[width=0.98\columnwidth]{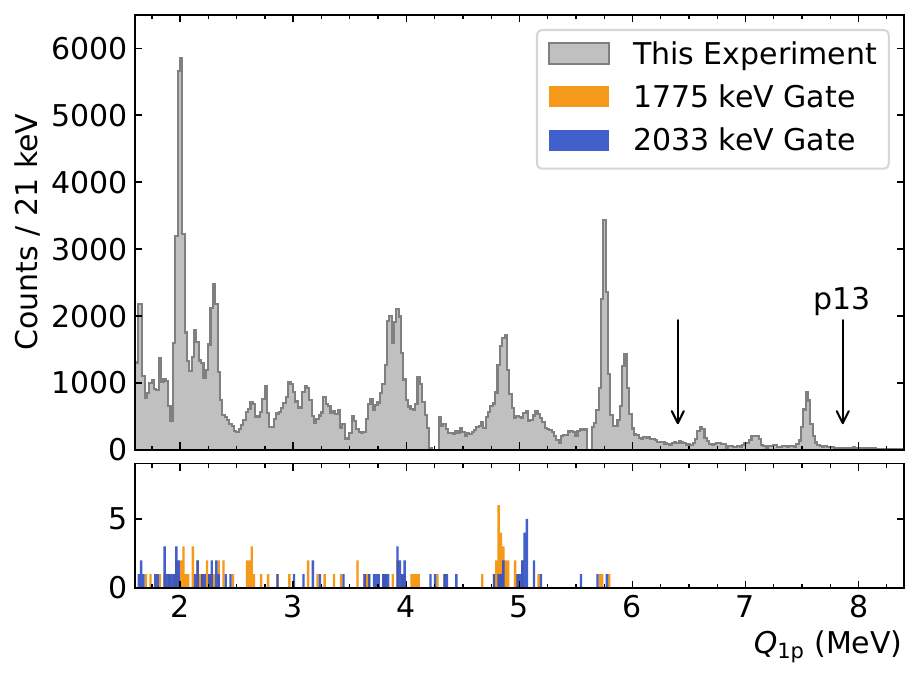} % The width is a tiny bit smaller compared to Q2p in order for them to have the same height.
\caption{\label{fig:q1p}%
 The measured one-proton decay energy, $Q_{1p}$, spectrum for $^{26}$P (top) with the same spectrum gated on specific coincident $\gamma$-rays (bottom). The 1775~keV $\gamma$-ray originates from the decay of the 2720~keV $7/2^+$ state in \alum{}, and the 2033~keV $\gamma$-ray from the 2485~keV $1/2^+$ state in \alum{}. Arrows indicate where the 11912~keV (left) and 13380~keV (right) states in $^{26}$Si reported by Liu \textit{et al.}~\cite{Liu2022} should be. Our spectrum shows no peak structure at the expected location of the previously purported 13380~keV state.}
\end{figure}

The solid angle coverage of each detector is not only dependent on the placement of the detectors, since the frame of the catcher foil shadows parts of the detectors \cite{Jensen2024Si25}.
Furthermore, the event reconstruction algorithm has to be taken into account:
An event with a small opening angle between the two protons may hit the same or neighboring strips in a detector, making the event difficult to reconstruct.
Monte Carlo-simulations, using the GEANT4 toolkit \cite{AGOSTINELLI2003250}, of detection within our setup of two-proton events with opening angles between 0 and 180\textdegree{} have been carried out to account for these effects.

\begin{figure}[t]
\centering
\includegraphics[width=0.99\columnwidth]{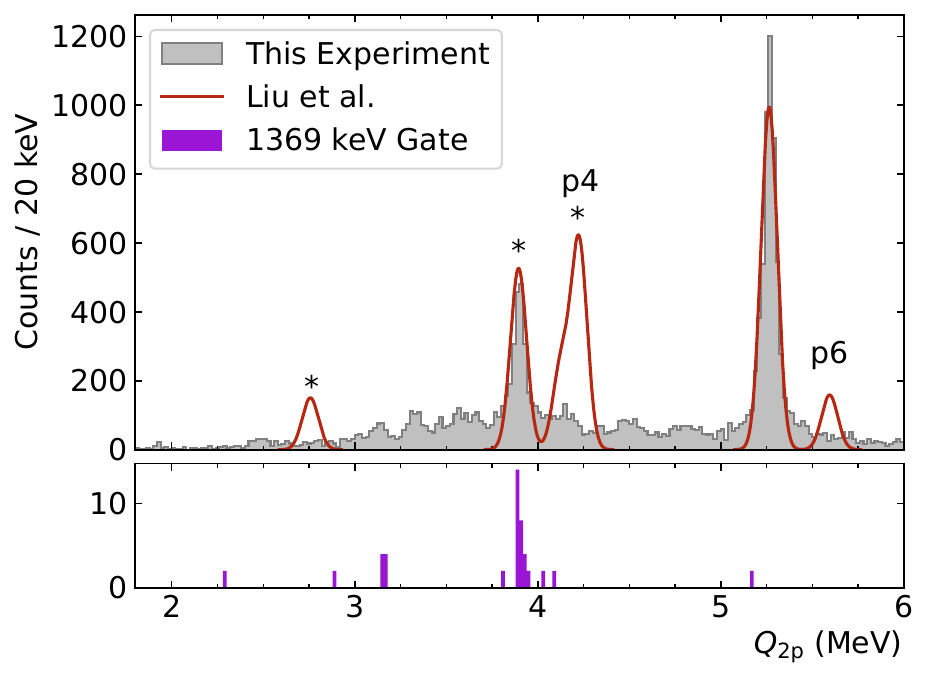}
\caption{\label{fig:q2p}%
The measured two-proton decay energy, $Q_{2p}$, spectrum for $^{26}$P compared to the distribution proposed in Ref.~\cite{Liu2022} (top). The solid grey line represents the experimental data from the present FRIB measurement. The solid red line denotes the expected contribution function, simulated using the decay energies and relative intensities for the 11912~keV and 13380~keV states in $^{26}$Si reported by Liu \textit{et al.}~\cite{Liu2022}. Decays to the first excited state in $^{24}$Mg are marked with an asterisk. The spectrum in the bottom shows the experimental data in coincidence with the \magn{} $2^+$ $\gamma$ energy. Our spectrum shows no peak structure for the reported 13380~keV state.}
\end{figure}

To accurately reconstruct the decay scheme, detailed kinematic event-by-event analyses were performed for both the $1p$ and $2p$ emission channels. The total one-proton and two-proton decay energies, $Q_{1p}$ and $Q_{2p}$, are determined by summing the measured kinetic energies of the emitted protons, $E_1$ and $E_2$, and the kinetic energy of the unobserved recoiling daughter nucleus ($^{25}$Al or $^{24}$Mg, respectively). Because the $^{26}$P parent decays at rest in the thin carbon foil, conservation of momentum, $\mathbf{p}$, dictates \mbox{$\mathbf{p}_{\mathrm{recoil}} = -\sum \mathbf{p}_p$}. For the $1p$ channel, the recoil energy reduces to \mbox{$E_{\mathrm{recoil}} = (m_p/M_{25})E_1$}, where $m_p$ is the proton mass and $M_{25}$ is the mass of the $^{25}$Al recoil. For the $2p$ channel, the recoil energy is dependent on the opening angle $\theta_{12}$ between the two emitted protons, which is determined from the segmentation of the silicon detector telescopes, \mbox{$E_{\mathrm{recoil}} = m_p \left( E_1 + E_2 + 2\sqrt{E_1 E_2} \cos(\theta_{12}) \right)/M_{24}$},
where $M_{24}$ is the mass of the $^{24}$Mg recoil. This kinematic reconstruction, coupled with the elimination of $\beta$-summing through the use of segmented, thin $\Delta E$ detectors, produces clean,  high-resolution $Q_{1p}$ and $Q_{2p}$ spectra including coincidence information with the $\gamma$-ray detectors.
$Q_{1p}$ and $Q_{2p}$ spectra are shown in Figs.~\ref{fig:q1p} and \ref{fig:q2p}, respectively.

Our $\beta$-delayed proton data contain numerous branches, both known and newly observed. The two main peaks from our $Q_{2p}$ spectrum are decays from the IAS to the first excited state and the ground state of \magn{} (Fig. \ref{fig:q2p}).
Overlays of the hypothetical contributions, estimated using the state energies and relative intensities published in Ref.~\cite{Liu2022}, reveal a clear mismatch with our measured $2p$ spectrum, hence our data challenge the existence of the newly proposed high-lying states in $^{26}$Si. The continuum in our spectrum originates from transitions to \magn{} fed by Gamow-Teller decays of \phos{} and shows no clear peak structures at the predicted $Q$-values of the reported transitions in Ref. \cite{Liu2022}. The case against the 13380~keV state is particularly compelling: The strongest evidence cited for it is a peak ($p_4$) at $Q_{2p} = 4250$~keV observed under a 1369~keV $\gamma$-ray coincidence gate (Fig. 1(d) in Ref.~\cite{Liu2022}).
This energy coincides with that of the expected $\beta$-delayed one-proton emission from the $^{25}$Si IAS to the $2^+$ state of \magn{}, a $\beta p \gamma$ sequence from a natural contaminant in projectile-fragmentation experiments that produces a prompt 1369~keV $\gamma$-ray, rendering it indistinguishable from a genuine $\beta 2p$ event under a $\gamma$ gate when the two-proton emission is detected only via the sum of the proton energies. No peak at this energy is present in our $Q_{2p}$ spectrum and no $\gamma$-rays are measured in coincidence with two protons at that energy.

The remaining peaks attributed to the 13380~keV state ($p_6$ at $Q_{2p}=5630$~keV and $p_{13}$ at $Q_{1p}=7854$~keV in Ref. \cite{Liu2022}) are also absent from our data; the $p_6$ energy region is accounted for by known $1p$ transitions in \sili{}. An arrow at $7854$~keV indicates where $p_{13}$ would be in our $1p$ spectrum in Fig. \ref{fig:q1p} where there is no clear peak structure.
Consequently, the claim of anomalous isospin mixing in the decay of \phos{} is refuted.

Having established the non-existence of these states from the spectral shapes alone, a combined evaluation of the absolute $\beta$-decay branching ratios for the $1p$ and $2p$ emissions from the IAS elucidates the origin of the previously reported Fermi strength fragmentation.
Table~\ref{tab:IAS_channels} provides information on nine identified decay channels of the IAS: Two 2-proton branches populating states in \magn{} (which each are composed of many individual channels) and seven 1-proton branches populating states in \alum{}, depicted in the decay scheme in Fig. \ref{fig:decayscheme}.
The seven 1-proton channels consist of transitions to the five bound states in \alum{} and to two unbound states with dominating $\gamma$-decay branches.
Feeding of the 2485~keV ($1/2^+$) state in \alum{} is identified here for the first time in $^{26}$P $\beta$-decay.
While feeding of the 2720~keV ($7/2^+$) state was previously identified via Doppler-shifted $\gamma$-rays by Schwartz \textit{et al.}~\cite{SCHWARTZ2015}, here we provide the first direct observation of the protons feeding this level.

The observed number of 1-proton and 2-proton decays of the IAS are converted into branching ratios by correcting for detection efficiencies and by normalizing to the absolute branching ratio determined in \cite{Thomas2004} for the 1983 keV one-proton peak (the most intense peak in our Fig. \ref{fig:q1p}).
This analysis yields an estimated total IAS $\beta$-feeding of $4.2(2)\%$, comprising a $3.2(2)\%$ $1p$ branch and a $0.99(12)\%$ $2p$ branch.
Compared to the branching ratios reported by Thomas {\it et al.}~\cite{Thomas2004} and by Liu {\it et al.}~\cite{Liu2022} there are substantial differences, see Table \ref{tab:IAS_channels}.
For the 2-proton transitions, we deduce significantly lower branching ratios.
Our analysis is based on 2-proton coincidences rather than 2-proton sum-spectra, which are susceptible to contributions from overlapping one-proton peaks.
For the transitions to the five bound states, the agreement with Thomas {\it et al.}~\cite{Thomas2004} is better.

The two-proton decay energies are measured with high precision from the combined detector array, giving $Q_{2p}$ values of $5267.7(0.9)$~keV and $3899.8(1.6)$~keV for the two branches. Using the two-proton separation energy $S_{2p}(^{26}\mathrm{Si}) = 7785.37(11)$~keV~\cite{AME2020}, these yield the IAS excitation energy in \sili{} as $E_x(\mathrm{IAS}) = 13052.9(1.0)$~keV and $13053.9(1.7)$~keV respectively. The weighted mean, $E_x(\mathrm{IAS}) = 13053(3)$~keV, including accounting for a systematic uncertainty on the calibration, is consistent with the value of $13055(2)$~keV reported by Liu~\textit{et al.}~\cite{Liu2022}. Independent determinations from the seven single-proton branches, obtained from fits to the $Q_{1p}$ spectra span the range 13051--13067~keV, in agreement with, but less precise than the two-proton result.

\begin{table}
\caption{\label{tab:IAS_channels}%
Absolute $\beta$-decay branching ratios for IAS channels in the decay of \phos{}, compared with previous measurements by Thomas {\it et al.}~\cite{Thomas2004} and Liu {\it et al.}~\cite{Liu2022}.
Two-proton ($2p$) channels feed the ground state and first excited state of \magn{}; one-proton ($1p$) channels feed levels in \alum{} \cite{Basunia2025}.
The 2485~keV ($1/2^+$) state in \alum{} ($^\dagger$) is identified for the first time in $^{26}$P $\beta$-decay. The 2720~keV ($7/2^+$) state ($^\ddagger$) was previously identified via Doppler-shifted $\gamma$-rays by Schwartz \textit{et al.}~\cite{SCHWARTZ2015}; here we provide the first direct proton-detection confirmation.
}
\begin{ruledtabular}
\begin{tabular}{lccc}
    Final state                & \multicolumn{3}{c}{Branching ratio (\%)}                           \\
    $J^\pi$, $E_x$ (keV)       & This work           & Thomas \textit{et al.} & Liu \textit{et al.} \\
\colrule
\multicolumn{4}{c}{$2p$ emission to \magn{}}                                                        \\
\colrule
$0^+$, 0                     & $0.75(8)$           & $1.4(2)$             & $1.2(2)$          \\
$2^+$, 1369                  & $0.24(9)$           & $0.79(12)$             & $0.6(2)$          \\
\quad \textit{Total} $2p$      & $\mathbf{0.99(12)}$ & $2.2(3)$             & $1.8(3)$          \\
\colrule
\multicolumn{4}{c}{$1p$ emission to \alum{}}                                                        \\
\colrule
$5/2^+$, 0                   & $0.41(3)$           & $0.61(5)$              & $0.29(4)$           \\
$1/2^+$, 452                 & $0.11(3)$           & $0.17(2)$              & $0.18(3)$           \\
$3/2^+$, 945                 & $0.24(2)$           & $0.21(8)$              & $0.12(2)$           \\
$7/2^+$, 1613                & $0.67(7)$           & $0.73(13)$             & $0.43(9)$           \\
$5/2^+$, 1790                & $1.27(11)$          & $1.40(11)$             & $0.81(14)$          \\
\quad \textit{Subtotal}        & $2.70(14)$          & $3.1(2)$             & $1.8(2)$          \\
$1/2^+$,\ 2485$^\dagger$     & $0.13(3)$           & ---                    & ---                 \\
$7/2^+$,\ 2720$^\ddagger$     & $0.38(7)$           & ---                    & ---                 \\
\quad \textit{Total} $1p$      & $\mathbf{3.2(2)}$ & $3.1(2)$             & $1.8(2)$          \\
\colrule
    \begin{tabular}{@{}l@{}} \quad \textit{Total IAS} \\ \quad \textit{$\beta$-feeding} \end{tabular}
                               & $\mathbf{4.2(2)}$ & $5.3(3)$             & $3.6(4)$
\end{tabular}
\end{ruledtabular}
\end{table}

The recently determined mass of \phos{} by Chen~\textit{et al.}~\cite{Chen2025} enables a more precise comparison of the total branching ratio to the IAS with the expectation from the superallowed Fermi sum rule. For a pure isospin \mbox{$T=2 \to T=2$} Fermi transition from \phos{}, the expected Fermi strength is \mbox{$B_F = T(T+1) - T_z(T_z+1) = 4$}, where \mbox{$T_z=-2$} is the projection of the isospin quantum number $T$. Using \mbox{$(ft)_0 = 6144(4)$~s}~\cite{Hardy2020}, the expected $ft$-value from the Fermi strength alone is \mbox{$ft = (ft)_0 / B_F = 1536(1)$~s}. Together with \mbox{$E_x(\mathrm{IAS}) = 13053(3)$~keV}, determined above, the electron capture $Q$-value from Chen~\textit{et al.} \cite{Chen2025}, \mbox{$Q_\mathrm{EC}(^{26}\mathrm{P}) = 18138(10)$~keV}, gives a phase space factor \mbox{$f =$~1362(15)} to the IAS, using the BetaShape code~\cite{Mougeot2019}. With the total half-life of \phos{} of \mbox{$T_{1/2} = 43.6(3)$~ms}~\cite{Liang2020} the expected branching ratio to the IAS from the Fermi strength alone is 3.87(5)\%.
With our observed total IAS $\beta$-feeding of 4.2(2)\%, the observed $\beta$-strength to the IAS is \mbox{$B=$ 4.3(2)}.
Hence, there is no indication from the branching ratios in
Table~\ref{tab:IAS_channels} that significant Fermi strength is missing in the feeding to the IAS.

Thomas {\it et al.} \cite{Thomas2004} deduce an IAS branching ratio significantly larger than ours. A Gamow-Teller feeding to the IAS of more than 25\% of the Fermi strength would be unusual. The absolute branching ratio to the IAS reported by Liu \textit{et al.}~\cite{Liu2022} is somewhat closer to our value.

Consequently, we demonstrate that the previously reported anomalous isospin mixing is an artifact of earlier experimental limitations. Our data restores agreement between experimental observations and standard expectations for isospin symmetry-breaking Fermi decay in the \emph{sd}-shell.

Beyond the findings of the present measurement, recent developments at FRIB have raised the available \phos{} rate to several thousand per second, two to three orders of magnitude above the 35-40 \phos{}/s of the present experiment, opening two avenues for more precisely quantifying isospin mixing in \sili{}. Firstly, the present analysis only determines the total branching ratio to the IAS without separating it into Fermi and Gamow-Teller components. The proton-$\beta$ angular correlation, accessible through the $\beta$-recoil energy shift of the proton peaks~\cite{THAYSEN1999194,FERNANDEZRUIZ2026140791}, would give direct sensitivity to the Fermi/Gamow-Teller composition of the IAS feeding and enable a proper separation of Fermi and Gamow-Teller strengths. Secondly, the particle decay of the $T=2$ IAS to the lower-$T$ states of \alum{} is isospin-suppressed, elevating the relative $\gamma$-decay branch of the IAS well beyond what is expected for a state at such high excitation energy. Observation of this branch, or a sensitive upper limit on it, would contribute to a more precise determination of the branching ratio to the IAS.

\begin{acknowledgments}

The authors acknowledge the support from the Independent Research Fund Denmark, Projects No. 2032-00066B, 4283-00172B, and 9040-00076B, from the Swedish Research Council, Project No. 2022-04248, from the U.S. National Science Foundation under Grants No. PHY-1913554, PHY-2209429, PHY-2514797, and PHY-2609824, from the Spanish MICIU/AEI/10.13039/501100011033, and from FEDER, EU, under Project No. PID2022-140162NB-I00.

This material is based upon work supported by the U.S. Department of Energy, Office of Science, Office of Nuclear Physics and used resources of the Facility for Rare Isotope Beams (FRIB) Operations, which is a DOE Office of Science User Facility under Award Number DE-SC0023633.

\end{acknowledgments}

\bibliography{main}

\end{document}